\documentclass[epsf,10pt]{article}
\usepackage{epsf}
\usepackage[T1]{fontenc}
\usepackage{amsmath}
\usepackage{amssymb}
\date{}

\title{Quantum mechanics in a cut Fock space}

\author{Maciej Trzetrzelewski \footnote{trzetrzelewski@th.if.uj.edu.pl} \\
M. Smoluchowski Institute of Physics, Jagiellonian University \\
Reymonta 4, 30-059 Cracow, Poland \\
  }
\begin{document}
\maketitle

\abstract{  A recently introduced numerical approach to quantum
systems is analyzed. The basis of a Fock space is restricted and
represented in an algebraic program. Convergence with increasing
size of basis is proved and the difference between discrete and
continuous spectrum is stressed. In particular a new scaling low
for nonlocalized states is obtained. Exact solutions for several
cases as well as general properties of the method are given. }

\section{Introduction}

Recently an attractive possibility  of modelling M-theory  through
relatively simple quantum mechanical systems [1] has occurred.
They emerge from the dimensional reduction of supersymmetric gauge
theories and provide a simple laboratory to study many properties
of supersymmetry [2,3] . It follows from [1] that there is an
strong connection between M-theory and $SU(N_c \longrightarrow
\infty )$ supersymetric Yang-Mills quantum mechanics (SYMQM).
However supersymmetric quantum mechanics have much longer history.
Various schemes have been analyzed to try to solve the hierarchy
problem including the idea of breaking SUSY. This was the reason
why SUSY was first studied in the simplest case of quantum
mechanics (SUSYQM)[2]. Apart from its physical meaning SUSYQM gave
also a deeper understanding of why certain potentials are
analytically solvable and other are not [4]. The SYMQM gauge
systems were studied for the first time in [3] where the exact
spectrum including the ground state of SYMQM D=2 was given. Later
on the extension for arbitrary SU(N) gauge group was also obtained
[5]. SUSYQM is known to have continuous spectrum due to the
fermion-boson cancellation [6]. According to BFSS hypothesis there
should be a bound state at  the threshold of the spectrum.
 However, since there are no exact solutions one is
forced to use numerical methods. \\

In this paper we  discuss in details a numerical approach of
solving quantum mechanical systems proposed in [7,8] and already
investigated in [9-13]. Next section contains  formulation of the
method as well as its general properties. We introduce a cutoff,
N, and by means of an algebraic program analyze a complete
dependence of the spectrum on the cutoff. We prove that the
eigenvalues converge towards exact (i.e in the infinite Hilbert
space) spectrum. In section 3 we give the exact spectrum of the
momentum and coordinate operators at arbitrary finite $N$. The
asymptotic behavior with $N \longrightarrow \infty$ is derived in
section 4 where a new scaling law, required to recover the
infinite Hilbert space limit, is formulated. The scaling  and its
universality is discussed in section 5 by giving the exact
spectrum of a free particle in quantum mechanics. Interestingly,
this solution differs only a little in comparison with the
eigenvalues of the hamiltonian for $D=2$ supersymmetric Yang-Mills
quantum mechanics at finite cutoff [14]. We prove that the
continuum spectrum in quantum mechanics gives rise to the
power-like dependance on the cutoff. This result is important in
studying supersymmetric systems where the distinction between
continuum and discrete spectra is an important issue. In section 6
we use numerical data in order to verify the theoretical results.
The implementation of the approach in Mathematica code will be
discussed there in details.

\section{A cut Fock space}

Every quantum hamiltonian can be represented in the eigenbasis of
a harmonic oscillator
\begin{equation} \{ \mid n \rangle = \frac{a^{\dagger n}}{\sqrt{n!}} \mid 0 \rangle, n \in N \},\end{equation}

\noindent where $a,a^{\dagger}$ are the normalized annihilation
and creation operators respectively. The  correspondence between
$a,a^{\dagger}$ and Q, P (coordinate and momentum operators
respectively) reads

\begin{equation}  Q=\frac{1}{\sqrt{2}}(a+a^{\dagger}),  P=\frac{1}{\sqrt{2i}}(a-a^{\dagger}).   \end{equation}

\noindent Since this basis is countable it is very convenient to
use it in numerical applications.  One can limit (1), e.g. $n
\leqslant N$, then calculate the finite matrix representation of
any hamiltonian and numerically diagonalize  above finite matrix
to obtain complete spectrum and the eigenstates of the system
\footnote{We are considering here hamiltonians with potentials
being polynomials in variable $Q$. Other types of potential
functions (e.g. $\frac{1}{r}$) may be analyzed as well by
introducing coordinate representation, however numerically it is
more time consuming.}. The procedure is simple and essentially
numerical, however a number of theoretical questions arises while
analyzing it. They will be discussed in this paper. \vspace{1cm}

 We denote \\
$H^{(N)}=[H]_{i,j}=h_{i,j} \ \  i,j=1,\ldots,N+1$ as operator $H$ in a cut Fock space (cutoff=N) , \\
$E^{(N)}_m$ and $ {c^{(N)}}_{m} = [{{c^{(N)}}}]^{j}_{m}$ \ where \
\ $j=1,\ldots,N+1$, $m=1,2, \ldots,N+1$ as eigenvalues and
eigenvectors of $H^{(N)}$ respectively, $E_m$ and $ c_m=[c_m ]^j $
\ \ as eigenvalues and eigevectors of $H$ respectively.
 In other words  \\

\begin{equation}
H^{(N)} {c^{(N)}}_m =E^{(N)}_m c^{(N)}_m \ \ \hbox{and} \ \
H c_m =E_m c_m.
\end{equation}

 The main aim of present work is to understand the dependence of
the spectrum of $H^{(N)}$ on $N$.
\newline

\section{ The spectrum of cut momentum and coordinate operators }

Matrix elements of the $P$ and $Q$ operators in the occupation
number basis read

 \begin{equation}    \langle n \mid Q \mid  k \rangle  =
 \sqrt{\frac{k}{2}}\delta_{n,k-1}+\sqrt{\frac{k+1}{2}}\delta_{n,k+1} \ \ , \ \
 \langle n \mid P \mid  k \rangle =
 \frac{1}{i}\sqrt{\frac{k}{2}}\delta_{n,k-1}+\frac{1}{i}\sqrt{\frac{k+1}{2}}\delta_{n,k+1}  .\end{equation}

\noindent In the Hilbert space limited to maximum of $N$ quanta
the eigenvalues of, e.g., momentum are given by zeros of
 the determinant

  \begin{equation}  I_{N+1} = \left\vert\begin{array}{cccccc}
        -\eta\    & \sqrt{1}  & 0        & .         & .        & .            \\
        -\sqrt{1} & -\eta     & \sqrt{2} & .         & .        & .             \\
        0         & -\sqrt{2} & -\eta    & .         & .        & .             \\
        .         & .         & .        & .         & .        & .            \\
        .         & .         & .        & .      & -\eta       & \sqrt{N}      \\
        .         & .         & .        & .      & -\sqrt{N} &-\eta

             \end{array}\right\vert ,
  \hbox{where} \ \  \eta=i\sqrt{2}\lambda . \end{equation}

\noindent Determinant (5) is evaluated by solving recursion
relation following from the Laplace expansion.  Making a change of
variables $ J_{N} = \frac{1}{N!} I_{N} $ we obtain

\begin{equation}  (N+2)J_{N+2} + \eta J_{N+1} - J_{N} = 0,\ \ J_{1}=-\eta,\ \ J_{2}=\frac{1}{2} ({\eta}^{2} + 1).
   \end{equation}

\noindent This recursion may be solved using the generating
function method. Let us define series with coefficients $J_{N}$

\begin{equation}  F(x)= J_{1} x + J_{2} x^{2} + J_{3} x^{3} + J_{4} x^{4} + \ldots  = \sum_{N=1}^{\infty} J_{N}x^{N}.
 \end{equation}

\noindent It follows from (6) that $F(x)$ satisfies

\begin{equation}(-x+\eta)(F(x)+1)=-F'(x),  \end{equation}

\noindent  with the boundary condition $F(0)=0$. The solution
reads

\begin{equation}
F(x)=exp \left( \frac{x^{2}}{2} - x\eta \right)-1 =
\sum_{N=1}^\infty \frac{H_{N}(i\frac{\eta}{\sqrt{2}})}{N!}
\left(i\frac{x}{\sqrt{2}}\right)^{N},
\end{equation}

\noindent where $H_{N}(x)$ stands for N-th Hermite polynomial.
Since $F(x)$ is analytic at $x=0$, the expansion (9) is
unambiguous so $J_{N}=2^{-\frac{N}{2}} i^{N}
\frac{H_{N}(i\frac{\eta}{\sqrt{2}})}{N!} $. Then

\begin{equation}  I_{N}=(-1)^{N} 2^{-\frac{N}{2}} i^{N} H_{N}(\lambda). \end{equation}

\noindent  It is clear now that the spectrum of operator $P$ in a
cut Fock space is given exactly by zeros  of Hermite polynomials.
Therefore, denoting  $p_{m}^{N}$ as the m-th eigenvalue of cut
momentum $P^{(N)}$, we  get

\begin{equation}   p^{N}_{m}=z^{(N+1)}_{m}, \ \ \ \   \hbox{where} \ \ \ \   H_{N+1}(z^{(N+1)}_{m})=0 ,
\ \ \ \ m=1,2,\ldots,N+1. \end{equation}

\noindent  This result will be used several times below. \\

  Calculation for coordinate operator $Q$ is very similar.
Recursion relation is slightly different but initial conditions
also change. Those two differences cancel each other and finally
we obtain the same result as for $P$. Therefore, denoting
$q_{m}^{N}$ as the m-th eigenvalue of cut coordinate $Q^{(N)}$, we
obtain

\begin{equation}   q^{N}_{m}=z^{(N+1)}_{m}, \ \ \ \   \hbox{where} \ \ \ \   H_{N+1}(z^{(N+1)}_{m})=0 ,
\ \ \ \ m=1,2,\ldots,N+1. \end{equation}

\noindent Since roots of Hermite polynomials are symmetric around
$0$,  we  consider only positive ones for which we introduce the
following enumeration

\begin{eqnarray}
0=p^{N}_{0} < p^{N}_{1} < p^{N}_{2} \ldots < p^{N}_{m} < \ldots < p^{N}_{\frac{N}{2}}  \ \ \hbox{, \ N - even},  \\
p^{N}_{1} < p^{N}_{2} \ldots < p^{N}_{m} < \ldots < p^{N}_{\frac{N+1}{2}} \ \ \hbox{, \ N - odd.}
\end{eqnarray}

\section{The continuum limit -- scaling}

Because of the continuum limit it is particularly interesting to
analyze the behavior of roots of  Hermite polynomials when $N \to
\infty$. It is possible to obtain the asymptotic relation (
details are in appendix  B )

\begin{eqnarray}
p_{m}^{N}=  \frac{\pi m}{\sqrt{2N+3}} \sqrt{ 1+\frac{{{\pi}^{2}}m^{2} -\frac{3}{2}}{3(2N+3)^{2}} }+O(N^{-4.5}),
 \ \ m=1,2\ldots,\ldots, \frac{N}{2} \ \  \hbox{N - even},  \\
p_{m}^{N}=  \frac{\pi(m-\frac{1}{2})}{\sqrt{2N+3}} \sqrt{ 1+\frac{{{\pi}^{2}}(m-\frac{1}{2})^{2}
-\frac{3}{2}}{3(2N+3)^{2}} }+O(N^{-4.5}), \ \ m=1,2\ldots,\ldots, \frac{N+1}{2} \ \ \hbox{N - odd}.
\end{eqnarray}

\noindent  If one naively evaluates the limit $N \to \infty$ for
fixed $m$ one  obtains $\underbrace{lim_{N \to \infty}}_{m \
fixed} p_{m}^{N}=0$. This is unacceptable, because we know that
$\underbrace{lim_{N \to \infty}}_{continuum} p_{m}^{N}=p$ with $p
\ne 0$ and $p \ne \infty$. It is clear now that $m$ has to depend
on $N$ as follows

\begin{equation} m=\frac{\sqrt{2N}}{\pi}p+b . \end{equation}

\noindent A  prescription, that guarantees existence of the
continuum limit

\begin{equation}  \lim_{N \longrightarrow \infty} p_{m(N)}^{N} = const.,  \end{equation}

\noindent is called scaling. The dependence (17) is universal,
that is for a large class of observables one obtains nontrivial
values when $N \to \infty$. Substituting (17) to (16) and ordering
resulting expression with respect to powers of $N$ we get

\begin{eqnarray} p_{m}^{N}= p + \frac{(b-\frac{1}{2})\pi}{\sqrt{2}} \frac{1}{\sqrt{N}} +
\frac{1}{12} p(p^2 -3)\frac{1}{N}+
\frac{\pi}{4\sqrt{2}}(b-\frac{1}{2})(p^2 -1)
\frac{1}{N^{\frac{3}{2}}}+\ldots \ \ .
\end{eqnarray}

\noindent Notice that $b$ has no influence on the result obtained
in the  continuum limit. Nevertheless, it is clear that taking
$b=\frac{1}{2}$ gives the best convergence.

\noindent  Moreover one can put

\begin{equation} m=\frac{\sqrt{2}}{\pi}p \sqrt{N}+b+\frac{c}{\sqrt{N}},    \end{equation}

\noindent providing another parameter that controls the
convergence. Now equation (19) is modified to

\begin{equation}    p_{m}^{N}= p + \frac{(b-\frac{1}{2})\pi}{\sqrt{2}} \frac{1}{\sqrt{N}} +
 \left(\frac{c\pi}{\sqrt{2}}+ \frac{1}{12} p(p^2 -3)\right) \frac{1}{N}
+ \frac{\pi}{4\sqrt{2}}(b-\frac{1}{2})(p^2 -1)
\frac{1}{N^{\frac{3}{2}}}+\ldots \ \ .\end{equation}

\noindent The procedure may by continued but the optimal values of
the coefficients $b,c$ are not universal, i.e. if we take another
observable they will be different. We can see this already in the
example (19) where coefficient $b$ depends on parity of $N$.
Nevertheless the limit (18) is valid for different observables.\\

 It is interesting to deal with the problem of cardinality of the
spectrum of the momentum operator. For all $N$ the spectrum of cut
operators consists of finite number of eigenvalues but we know
that in the continuum limit there has to be an uncountable set of
eigenvalues. How those two facts can be brought together?
According to (15,16), for large N, there are $N$ eigenvalues
($\frac{N}{2}$ positive ones and $\frac{N}{2}$ negative ones, or
$\frac{N-1}{2}$ positive ones and $\frac{N-1}{2}$ negative ones
for $N$ even or odd respectively) separated by the distance
$O(\frac{1}{\sqrt{N}})$. It means that the spectrum becomes denser
so that it is possible to chose such $m(N)$ that

\begin{equation}  \forall_{p \in R}   : lim_{N \to \infty} p_{m(N)}^{N}=p .\end{equation}

\noindent Therefore the set of all roots of all Hermite
polynomials $\mathcal{Z}=\bigcup_{N=1}^{\infty} spectrum(
\hat{P}^{(N)} )$ is dense in $\mathbb{R}$. However it is not equal
to $\mathbb{R}$ because $\mathcal{Z}$ is countable due to the fact
that there is countable amount of Hermite polynomials. In other
words elements of $\mathcal{Z}$ behave similarly to rational
numbers in $\mathbb{R}$. It looks as if there was something wrong
because the spectrum of operator $P$ should be continuous. In
order to solve this paradox we use the scaling  (17). Now any real
number $p$ can be obtained in the continuum limit so that all
elements of $\mathbb{R}$ are reproduced.

\section{ The spectrum of the cut kinetic energy }

In order to calculate the eigenvalues of a free particle  we
introduce the cut parity operator

\begin{equation}
 {\Sigma}^{(N)}  = \underbrace{\left[ \begin{array}{ccccccc}
        1  &  0   &           & 0            \\
        0  & -1   &           &             \\
           &      & \ddots    &             \\
        0  &      &           & (-1)^N
          \end{array}\right] }_{N+1 \ \ \hbox{columns}}.
   \end{equation}

\noindent A straightforward calculation shows that $ [ (p^2)^{(N)}
, {\Sigma}^{(N)}  ]=[  (p^{(N)})^{2}  , {\Sigma}^{(N)}  ] =0$ so
that
 $ (p^2)^{(N)} $ as well as $(p^{(N)})^{2}  $, represented in an eigenbasis of
 ${\Sigma}^{(N)}$, splits into two blocks.

\noindent Let $N$ be an odd number. In this case the matrices
$(p^2)^{(N)}$ and $(p^{(N)})^{2}  $    contain  two blocks
 ${\frac{N+1}{2} }\times { \frac{N+1}{2} }$ each. We have
\footnote{The dimensions of those blocks are equal in this case
because for odd $N$ the
 rank of the matrix ${\Sigma}^{(N)}$ is  $N+1$ - even. Therefore it contains the same number of $+1$ and $-1$.   }

\begin{equation}
 H^{(N)}=(\frac{1}{2}  {p^2})^{(N)}  =
\left[ \begin{array}{cccccc}
                 & A^{\frac{N+1}{2}}_{+}  &   &   \multicolumn{1}{|c}{}       &       0                 &     \\
 \cline{1-6}
                 &      0                 &   &   \multicolumn{1}{|c}{}       &  A^{\frac{N+1}{2}}_{-}  &     \\
          \end{array}\right] ,
\frac{1}{2}   (p^{(N)})^{2}  =
\left[ \begin{array}{cccccc}
                 & B^{\frac{N+1}{2}}_{+}  &   &   \multicolumn{1}{|c}{}       &       0                 &     \\
 \cline{1-6}
                 &      0                 &   &   \multicolumn{1}{|c}{}       &  B^{\frac{N+1}{2}}_{-}  &     \\
          \end{array}\right] ,
\end{equation}
 \noindent where

\begin{equation}
   A^{(M)}_{+}    =
   \left[ \begin{array}{ccccccc}
   \frac{1}{4}             &  -\frac{\sqrt{1\cdot 2}}{4} &              & 0        \\
-\frac{\sqrt{1\cdot 2}}{4} & \frac{5}{4}              & \ddots          &       \\
                           & \ddots                   & \frac{4M-7}{4}  &  -\frac{\sqrt{(2M-3)\cdot (2M-2)}}{4} \\
        0                  &                          &  -\frac{\sqrt{(2M-3)\cdot  (2M-2)}}{4} & \frac{4M-3}{4}        \end{array}\right] ,
 \end{equation}

\vspace{1cm}

\begin{equation}
  A^{(M)}_{-}    =
   \left[ \begin{array}{ccccccc}
   \frac{3}{4}             &  -\frac{\sqrt{2\cdot 3}}{4} &              & 0        \\
-\frac{\sqrt{2\cdot 3}}{4} & \frac{7}{4}              & \ddots          &       \\
                           & \ddots                   & \frac{4M-5}{4}  &  -\frac{\sqrt{(2M-2)\cdot (2M-1)}}{4} \\
        0                  &                          &  -\frac{\sqrt{(2M-2)\cdot  (2M-1)}}{4} & \frac{4M-1}{4}        \end{array}\right] ,
   \end{equation}

\vspace{0.5cm}

\noindent and

\vspace{0.5cm}

\begin{equation}
   B^{(M)}_{+}    =
   \left[ \begin{array}{ccccccc}
   \frac{1}{4}             &  -\frac{\sqrt{1\cdot 2}}{4} &              & 0        \\
-\frac{\sqrt{1\cdot 2}}{4} & \frac{5}{4}              & \ddots          &       \\
                           & \ddots                   & \frac{4M-7}{4}  &  -\frac{\sqrt{(2M-3)\cdot (2M-2)}}{4} \\
        0                  &                          &  -\frac{\sqrt{(2M-3)\cdot  (2M-2)}}{4} & \frac{4M-3}{4}        \end{array}\right] ,
   \end{equation}

\vspace{1cm}

\begin{equation}
   B^{(M)}_{-}    =
   \left[ \begin{array}{ccccccc}
   \frac{3}{4}             &  -\frac{\sqrt{2\cdot 3}}{4} &              & 0        \\
-\frac{\sqrt{2\cdot 3}}{4} & \frac{7}{4}              & \ddots          &       \\
                           & \ddots                   & \frac{4M-5}{4}  &  -\frac{\sqrt{(2M-2)\cdot (2M-1)}}{4} \\
        0                  &                          &  -\frac{\sqrt{(2M-2)\cdot  (2M-1)}}{4} & \frac{4M-1}{4}-\frac{2M}{4}        \end{array}\right] .
   \end{equation}

\noindent Since $ A^{(M)}_{+}=B^{(M)}_{+}  $ (the + sign corresponds to odd m) therefore

\begin{eqnarray}
E^{N}_{m} \equiv (\frac{1}{2}
p^2)^{N}_{m}=\frac{1}{2}(p^{N}_{m})^2 \ \ \hbox{ where}  \ \
m=1,3,\ldots,N \ \  \hbox{for} \ \ N \ \ - \ \ \hbox{odd}.
\end{eqnarray}

\noindent When $N$ is  even, the analogous procedure gives

\begin{eqnarray}
 E^{N}_{m} \equiv (\frac{1}{2}  p^2)^{N}_{m}=\frac{1}{2}  (p^{N}_{m})^2 \ \  \hbox{ where}  \ \ m=2,4,\ldots,N \ \  \hbox{for} \ \ N \ \ - \ \  \hbox{even}.
 \end{eqnarray}

\noindent  Now it is useful  to present the eigenvalues (29,30) in
the following table. For example for $N<4$, $m<5$

\begin{center}
\setlength{\unitlength}{.3mm}
\begin{picture}(350,50)(40,-60)

\put(70,-65){\vector(1,0){300}}
\put(110,-50){\makebox(0,0){$N=0$}}
\put(170,-50){\makebox(0,0){$N=1$}}
\put(240,-50){\makebox(0,0){$N=2$}}
\put(300,-50){\makebox(0,0){$N=3$}}

\put(70,-65){\vector(0,-1){120}}
\put(40,-90){\makebox(0,0){$m=1$}}
\put(40,-115){\makebox(0,0){$m=2$}}
\put(40,-140){\makebox(0,0){$m=3$}}
\put(40,-167){\makebox(0,0){$m=4$}}

\end{picture}
\end{center}

\begin{equation}
   \left(\begin{array}{cccccc}

E^{0}_{1} ={\large\textbf{?}} & E^{1}_{1} = \frac{1}{2}(p^{1}_{1})^2  & E^{2}_{1} ={\large\textbf{?}}
& E^{3}_{1} = \frac{1}{2}(p^{3}_{1})^2     \\

   &         &           &                                                             \\

 -   & E^{1}_{2} ={\large\textbf{?}}     & E^{2}_{2} = \frac{1}{2}(p^{2}_{2})^2
     & E^{3}_{2} ={\large\textbf{?}}     \\

 &          &          &            \\

  -   & -  & E^{2}_{3} ={\large\textbf{?}}  & E^{3}_{3} = \frac{1}{2}(p^{3}_{3})^2
        \\

   &         &       &              \\

  -    & -    & -   & E^{3}_{4} ={\large\textbf{?}}    \\
 &      &    &            \\

   \end{array}\right).
   \end{equation}

\noindent We prove that fields filled with questionmarks in (31)
are equal to their neighbors on the right.

\noindent Let $N$ be an odd number. We have already shown that
matrix $(\frac{1}{2}p^2)^{(N)}$   contains two blocks (24). Now,
if we increase $N \longrightarrow N+1$ we obtain

\begin{equation}
 (\frac{1}{2}  {p^2})^{(N+1)}  =\left[ \begin{array}{ccccccc}
                 &                              &          &  \multicolumn{1}{|c}{\vdots}   & \multicolumn{1}{|c}{}       &               &    \\
                 &  A^{\frac{N+1}{2}}_{+}                 &          &  \multicolumn{1}{|c}{0}        & \multicolumn{1}{|c}{}       &       0       &     \\
                 &                              &          &  \multicolumn{1}{|c}{\bullet}  & \multicolumn{1}{|c}{}       &               &     \\  \cline{1-3}
               0 &       \ldots  \ \  \     0   &  \bullet &   0                            & \multicolumn{1}{|c}{}       &               &     \\  \cline{1-6}
                 &                              &          &                                & \multicolumn{1}{|c}{}       &               &     \\
                 &      0                       &          &                                & \multicolumn{1}{|c}{}       &  A^{\frac{N+1}{2}}_{-}  &     \\
                 &                              &          &                                & \multicolumn{1}{|c}{}       &               &
          \end{array}\right],
\end{equation}

\noindent so that the  block $A^{\frac{N+1}{2}}_{-}$ does not
change. \footnote{The matrix $A^{\frac{N+1}{2}}$ becomes larger
because the increment $N \longrightarrow N+1$ produces new state
with parity $\Sigma= +1$. Next increment (i.e. $N+1
\longrightarrow N+2$) will produce new state with parity
$\Sigma=-1$ ect.} It means that in the $N+1$ cutoff, eigenvalues
from this block remain untouched. This block corresponds to even
$m$, therefore we have

 \begin{equation}   E^{(N+1)}_{m} =E^{(N)}_{m}  \ \  \hbox{for} \ \ m=2,4,\ldots,N+1 \ \  \hbox{ N - odd}.
    \end{equation}

\noindent When $N$ is  even  an analogous procedure gives

\begin{equation}    E^{(N+1)}_{m} =E^{(N)}_{m}  \ \  \hbox{for} \ \ m=1,3,\ldots,N \ \  \hbox{ N - even} .
 \end{equation}

\noindent  That completes the whole spectrum of ${(\frac{1}{2}
p^2)}^{(N)}$. First few exemplary values are

\[
\left\vert\begin{array}{cccccc}
         0.25      & 0.25      & 0.137    & 0.137     & 0.095    & 0.095      \\
         -         & 0.75      & 0.75     & 0.459     & 0.459    & 0.333      \\
         -         & -         & 1.362    & 1.362     & 0.892    & 0.892      \\
         -         & -         & -        & 2.040     & 2.040    & 1.400      \\
         -         & -         & -        & -         & 2.762    & 2.762      \\
         -         & -         & -        & -         & -        & 3.516
  \end{array}\right\vert
.\]

\noindent Above numbers were obtained  from a program described in
next section and indeed confirm (33,34). According  to (15,16)
formulas (29,30,33,34) give

\begin{equation}
E^{(N)}_{m} \approx \frac{\pi^2}{2} \frac{(m-\frac{1}{2})^2}{2N+3}  \ \ \ \hbox{ N - odd} \ \ \ \hbox{ m - odd },\\
\end{equation}

\begin{equation}
E^{(N)}_{m} \approx \frac{\pi^2}{2} \frac{m^2}{2N+5}  \ \ \ \hbox{ N - odd} \ \ \ \hbox{ m - even },\\
\end{equation}

\begin{equation}
E^{(N)}_{m} \approx \frac{\pi^2}{2} \frac{(m-\frac{1}{2})^2}{2N+5}  \ \ \ \hbox{ N - even} \ \ \ \hbox{ m - odd },\\
\end{equation}

\begin{equation}
E^{(N)}_{m} \approx \frac{\pi^2}{2} \frac{m^2}{2N+3}  \ \ \ \hbox{ N - even} \ \ \ \hbox{ m - even },\\
\end{equation}

\noindent Note that (17) applied to each (35,36,37,38) separately
gives the  expected limit  $\frac{p^2}{2}$. Moreover we see that
the dependence of spectrum on $N$ is  power like i.e. slow.

\section{Applications}

In this section we  use above analytic  results to verify the
method introduced in [7,8]. It consists of numerical
diagonalization of finite matrices and extrapolation of results to
$N \to \infty$. Practically, when one deals with fast convergence
of eigenvalues it is sufficient to stop the calculations for
relatively low cutoff N ( in the case of one dimensional
nonrelativistic quantum mechanics the results for N=50 are already
very accurate ). Nevertheless  a problem may occur when  the
convergence is slow ( polynomial ), or when numerical calculations
are time consuming
even for low N.\\

 One of aims of this work is better understanding the case of a
free particle which has the former feature. The later situation
occurs  everytime when there are higher dimensions. Models
discussed in  [7,8] have both of those difficulties, therefore it
is crucial to  understand analytically the asymptotics of the
spectrum for large $N$. We expect that the power-like behavior in
$N$ is characteristic not only for the spectrum of a free particle
but also it occurs in every scattering problem because in those
cases the asymptotics of  wave functions is the same as for a free
particle so that the asymptotic momentum may be properly defined.

\subsection{Quantum mechanics on a computer}

Let us discuss in details the implementation of the method [7,8]
in the computer code. Consider quantum system with $D$ degrees of
freedom with $D$ creation and annihilation operators. One can
construct the whole orthogonal basis from the vacuum state $ \mid
0 \rangle $

\begin{equation}   \mid n_1,n_2, \ldots ,n_D  \rangle =
\frac{ (\widehat{a}_1^{ \dagger })^{n_1} }{\sqrt{n_1!}}
\frac{ (\widehat{a}_2^{ \dagger })^{n_2} }{\sqrt{n_2!}} \ldots
\frac{ (\widehat{a}_D^{ \dagger })^{n_D} }{\sqrt{n_D!}}  \mid 0 \rangle. \end{equation}

\noindent Each state in a cut Fock space, decomposed in this
basis, is represented as a list in Mathematica program

\[ \mid \psi \rangle =\sum_{k=1}^{p} a_k \mid n_1^k,n_2^k, \ldots ,n_D^k  \rangle  \longrightarrow  \]

\begin{equation} \Bigl\{ p,\{a_1,a_2, \ldots ,a_p \},
\bigl\{ \{n_1^1,n_2^1,\ldots,n_D^1 \},\{n_1^2,n_2^2,\ldots,n_D^2
\},\ldots,\{n_1^p,n_2^p,\ldots,n_D^p \} \bigr\} \Bigr\} .
\end{equation}

\noindent  The first element of this list specifies the number of
basis vectors, that the state $ \mid \psi \rangle$ is decomposed
on . The second element of the list is a list of coefficients of
this decomposition. Basis vectors are represented in the third
element of this list. For example

 $$a\mid 0,1 \rangle+ b\mid 1,0 \rangle +c\mid 1,1 \rangle \longrightarrow
  \Bigl\{ 3,\{a,b,c\}, \bigl\{ \{ 0,1 \},\{ 1,0 \},\{ 1,1 \} \bigr\} \Bigr\}. $$

\noindent The creation and annihilation operators

\begin{equation} \widehat{a}_k: \ \ \ \widehat{a}_k  \mid n_1,n_2, \ldots,n_k,\ldots ,n_D  \rangle  =
 \sqrt{n_k} \mid n_1,n_2, \ldots,n_k -1,\ldots ,n_D  \rangle , \end{equation}

\begin{equation} \widehat{a}_k^{ \dagger }: \ \ \
\widehat{a}_k^{ \dagger } \mid n_1,n_2, \ldots,n_k,\ldots ,n_D  \rangle  =
\sqrt{n_k +1} \mid n_1,n_2, \ldots,n_k +1,\ldots ,n_D \rangle .\end{equation}

\noindent have the following action in the list representation

$$  \widehat{a}_k \mid \psi \rangle \longrightarrow
\Bigl\{ p,\{\sqrt{n_k^1 }a_1, \ldots ,\sqrt{n_k^p }a_n \},
\bigl\{ \{n_1^1,\ldots,n_k^1 -1,\ldots,n_D^1 \},\{n_1^2,\ldots,n_k^2 -1,\ldots,n_D^2 \},\ldots,  $$

\begin{equation}  \ldots, \{n_1^p,\ldots,n_k^p -1,\ldots,n_D^p \} \bigr\} \Bigr\} , \end{equation}

\noindent and

$$ \widehat{a}_k^{ \dagger } \mid \psi \rangle \longrightarrow
\Bigl\{ p,\{\sqrt{n_k^1 +1 }a_1, \ldots ,\sqrt{n_k^p +1}a_p \},
\bigl\{ \{n_1^1,\ldots,n_k^1 +1,\ldots,n_D^1 \},\{n_1^2,\ldots,n_k^2 +1,\ldots,n_D^2 \},\ldots, $$

\begin{equation} \ldots \{n_1^p,\ldots,n_k^p +1,\ldots,n_D^p \} \bigr\} \Bigr\}   . \end{equation}

\noindent In order to evaluate the matrix representation of any
observble we define procedures which add and multiply on arbitrary
state by a complex number as well as scalar multiply states. For
example

$$\mid \psi \rangle \longrightarrow  \Bigl\{2,\{1,2\},\bigl\{ \{0,0\},\{0,1\} \bigl\} \Bigl\},$$

$$\mid \phi \rangle \longrightarrow  \Bigl\{2,\{1,1\},\bigl\{ \{0,2\},\{0,1\} \bigl\} \Bigl\},$$

\noindent then

$$\mid \psi \rangle +\mid \phi \rangle  \longrightarrow
\Bigl\{3,\{1,3,1\},\bigl\{ \{0,0\},\{0,1\},\{0,2\} \bigl\} \Bigl\},$$

$$ 2 \mid \phi \rangle \longrightarrow  \Bigl\{2,\{2,2\},\bigl\{ \{0,2\},\{0,1\} \bigl\} \Bigl\} ,   $$

\noindent and

$$ \langle \psi \mid \phi \rangle \longrightarrow   2 .     $$

\noindent Adding lists is simply adding those coefficients of the
decomposition (40),
 that have the same basis vectors. If decompositions of  $\mid \psi \rangle$ and $\mid \psi \rangle$
  have different basis vectors then the sublist consisting of basis vectors  has to be extended accordingly. \\

\noindent The procedure of multiplying the state by a number
reduces to multiplying the list of coefficients by this number. \\

\noindent Scalar multiplication $\langle \psi \mid \phi \rangle$
reduces to a search for common basis vectors occurring in
decomposition of $\mid \psi \rangle$ and $\mid \phi \rangle$.
Afterwards
proper coefficients and their complex conjugations have to be multiplied. \\

\noindent These rules  allow to automatically represent any
operator in a cut basis (39).

\subsection{Numerical diagonalization}

Here we compare numerical data and analytic results of section
3for

\noindent a1) eigenvalues of  $P^{(N)}$ evaluated by the program
described in 6.1. ( according to section 3 they are exactly the
roots of Hermite polynomials ),

\noindent a2)  the asymptotic form (16),

\begin{figure}[h]
\centering \leavevmode \epsfverbosetrue \epsffile{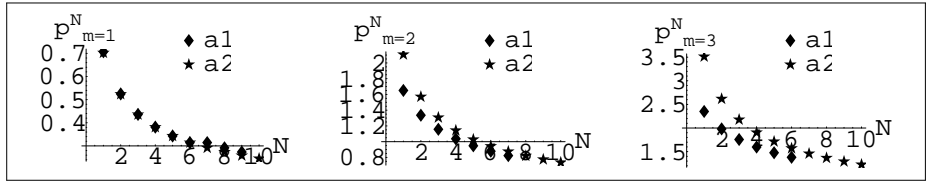}
\caption{Asymptotic, a2), and exact, a1), behavior of $ p^{(N)}_m$
for m=1,2,3.}
\end{figure}

\noindent Figure 1 presents the comparison of cases a1 and a2  for
m=1,2,3. The approximate value is obtained from (16) by taking
only the leading term

\begin{equation} p^{N}_{m} \approx \frac{\pi (m-\frac{1}{2})}{\sqrt{2N+1}}  .\end{equation}

\noindent We see that there is a good agreement between exact and
approximate values even for low $N$, and it gets worse for higher
$m$ where next terms of the expansion of (16) are important.

\subsection{Continuum limit on a computer}

\noindent Here we want to obtain dispersion relation that is the
dependence of the energy on momentum $E(p)$. Obviously we know
that $E(p)=\frac{p^{2}}{2}$ but it is only because we are able to
solve Schr\"{o}dinger equation for a free particle. However one
has to put himself in a situation where there is a certain set of
eigenvalues ${E}^{N}_{m}$ and no information about the dispersion
relation is available. In other words the question is how to
obtain unknown {\em a priori} function $E(p)$  by means of
eigenvalues ${E}^{N}_{m}$? In order to do this one has to make $m$
dependent on $N$ : $ m=m(N,p) $ such that the limit

\begin{equation} \lim_{N \to \infty} {E}^{N}_{m(N,p)} = E(p),  \end{equation}

\noindent is not trivial that is $E(p)< \infty$ and $E(p) \ne 0$.
Note that (46) automatically
 requires the set $\{ {E}^{N}_{m}: m,N\in \mathbb{N}\}$ to be dense in $E(\mathbb{R})$ .
In case of a free particle ($E(\mathbb{R})=[0,\infty)$) we can
even construct this set (squares of roots of Hermite polynomials)
however it is a general property of any operator with continuous
spectrum. This is exactly the reason why ${E}^{N}_{m}$ depend on N
as a power rather then exponentially.\\

 Let us emphasize that we don't have to know the dependence
$E(p)$ to  evaluate  $m(N)$. This is because the relation  $m(N)$
was established on grounds of the condition that there has to
exist the continuum limit  for the momentum, so that any other
operator commuting with $P$ will have the same scaling. We will
analyze in details the case of a free particle in nonrelativistic
quantum mechanics but another example may be Dirac equation where
we expect that the scaling law (18) will give
$E(p)=\sqrt{M^{2}+p^{2}}$. Therefore the scaling in (46) has to be
the same as for momentum operator, that is

\begin{equation} m(N,p) = \frac{\sqrt{2N}}{\pi}p  +\frac{1}{2} .\end{equation}

\noindent However in formula  (47) we have to introduce a certain
change

\begin{equation} m(N,p) \longrightarrow 2m(N,p)= 2 \frac{\sqrt{2N}}{\pi}p  +1,  \end{equation}

\noindent because the scaling (48) is  meant for positive
eigenvalues of operator $P^{(N)}$ only. Let us consider an example
of $N=7$. The spectrum of operator $P^{(8)}$ consists of roots of
$H_{8}(x)$, so that we have 8 roots where 4 of them are positive
and 4 are negative.

\begin{center}

\setlength{\unitlength}{.3mm}
\begin{picture}(500,50)(-100,-20)

\put(0,0){\vector(1,0){300}}

\put(0,0){\makebox(0,0){$*$}}
\put(40,0){\makebox(0,0){$*$}}
\put(80,0){\makebox(0,0){$*$}}
\put(120,0){\makebox(0,0){$*$}}
\put(140,0){\makebox(0,0){$0$}}
\put(160,0){\circle*{3.7}}
\put(200,0){\circle*{3.7}}
\put(240,0){\circle*{3.7}}
\put(280,0){\circle*{3.7}}

\end{picture}
\end{center}

\noindent Now, if we square them the spectrum becomes positive and
the numeration of eigenvalues changes as follows. \footnote{Since
roots of $H_n(x)$ are symmetric around the origin, their squares
will give double degeneracy. Hance for the free particle dots and
stars should be on the same point however in general it is not the
case. }

\begin{center}
\setlength{\unitlength}{.3mm}
\begin{picture}(500,50)(-100,-20)
\put(0,0){\vector(1,0){330}}

\put(140,0){\makebox(0,0){$0$}}
\put(180,0){\makebox(0,0){$*$}}
\put(200,0){\circle*{3.7}}
\put(220,0){\makebox(0,0){$*$}}
\put(240,0){\circle*{3.7}}
\put(260,0){\makebox(0,0){$*$}}
\put(280,0){\circle*{3.7}}
\put(300,0){\makebox(0,0){$*$}}
\put(320,0){\circle*{3.7}}

\end{picture}
\end{center}

\noindent For example, the eigenvalue  that we used to number as
the first one will now have the index $m=2$, the eigenvalue that
we used to number as the second one will now have the index $m=4$
ect. Therefore the formula (47) has to be rescaled as in (48).\\

According to  (48), eigenvalues  ${E}^{N}_{m}$ are analyzed by
fixing any momentum value $p$ and writing down the value
${E}^{N_{max}}_{m(N_{max},p)}$ where $N_{max}$  is the highest N
in computer calculations. Then we change the momentum value and
repeat the procedure. In this way one obtains an approximate (
because of limited value of $N$ ) dependence $E(p)$, which should
reproduce $\frac{p^2}{2}$  for a free particle. However the
problem concerning the formula $m(N,p)$ occurs because $m$ is not
a natural number. We circumvent  this by taking an integer part (
INT ) of Eq.(48), so that the matrix index is $INT(m(N_{max},p))$,
where $N$ is an even number. The convergence of those elements was
checked in Mathematica for $p= 1,2,3, \ldots , 37$ (e.g. Figure 2)

\begin{figure}[h]
 \centering \leavevmode \epsfverbosetrue \epsffile{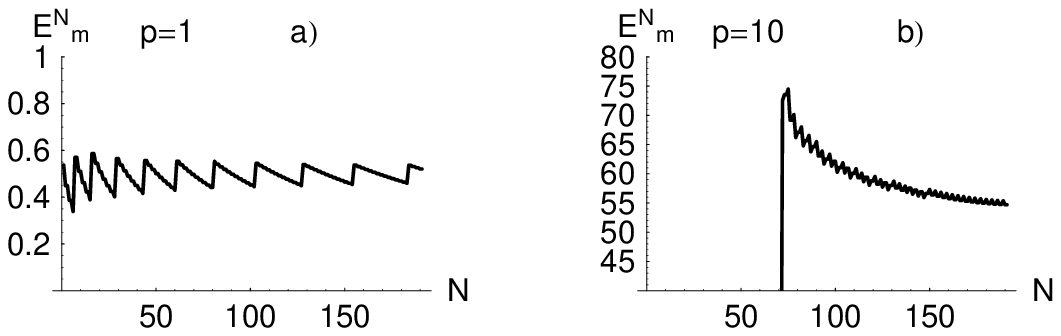}
\caption{The convergence of ${{E}}^{N}_{INT(m(N_{max},p))}$ for
p=1 and p=10 respectively.}

\centering \leavevmode \epsfverbosetrue \epsffile{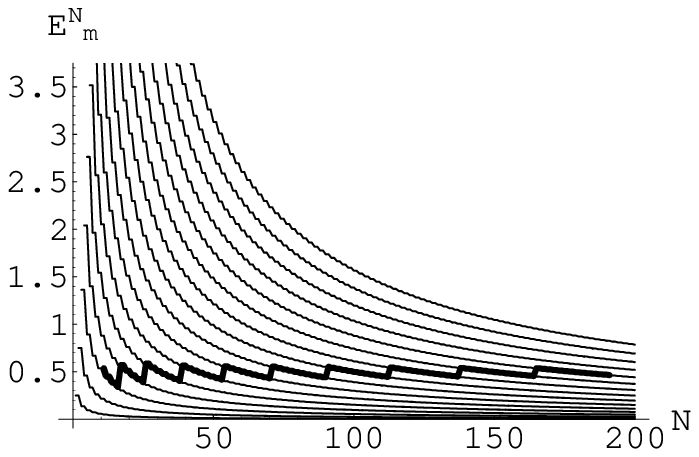}
\caption{The convergence of ${{E}}^{N}_{INT(m(N_{max},p))}$ for
p=1.}
\end{figure}

 This behavior can be understood as follows. If
one plots the dependence of ${E}^{N}_{1}$ on N ( p is fixed ), one
obtains (35-38) a hyperbola. The lower index $m=1$ specifies the
first eigenvalue. The upper index enumerates the cutoff. If we
plot the dependence of ${E}^{N}_{2}$ on N, we get another
hyperbola ect. Finally the plot of ${E}^{N}_{m}$ is a set of
hyperbolas on a plane ( see Figure 3). The scaling that we have
used previously means that from each hyperbola we are taking only
one point in such way that in the limit of large $N$ a constant
value is reproduced. Why on those figures we see cut hyperbolas
instead of points? This is because we had to introduce the INT
procedure which is equivocal. In a consequence it is possible that
for different cutoffs ( say N and N' ) there is $
INT(m(N,p))=INT(m(N',p) ) $. It means that points $(N , INT(m(N,p)
) )$ and $(N',INT(m(N',p)))$ are on the same hyperbola. Eventually
$N$ will be large enough so that the INT operation notices the
difference and the point "jumps" to next hyperbola. Let us also
note that the scaling (18) is an asymptotic law hance for low N
the behavior of ${E}^{N}_{m(N,p)}$ may vary for different values
of p. This effect
accounts for the different behavior in Figure 2a and 2b.     \\

 The dispersion relation  extracted in this way is presented in
Figure 4. This result has no error because all eigenvalues are
precisely evaluated hance any statistical interpretation is
meaningless. The tangent coefficient $0.5365$ is different from
$0.5$ but we did expect that because it is a numerical result
obtained on grounds of limited cutoff. Moreover we had to
introduce the INT operation. In a consequence we had to choose
only one point from cut hyperbolas. It is a source of a new error which gets smaller while the cutoff increases. \\

Therefore Figure 4 confirms that  we can obtain the
 dispersion relation from the knowledge of the spectrum of a cut hamiltonian.  \\

\begin{figure}[h]
\centering \leavevmode \epsfverbosetrue \epsffile{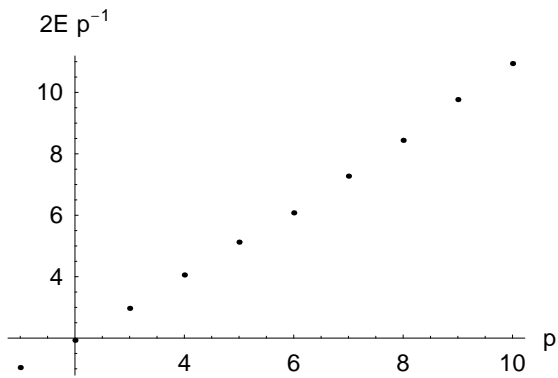}
\caption{Reproduced dispersion relation.  }
\end{figure}

\section{Bound states versus scattering states}

\noindent In this section we  stress the difference between
localized and nonlocalized states. It follows from simple algebra
( see Appendix A) that \footnote{The notation is explained in
section 2}

\begin{equation}
 E_m-E^{(N)}_m =
\frac{\sum_{j=1}^{N} \sum_{j=N+1}^{\infty} h_{i j} c^{j}_{m}
}{\sum_{i=1}^{N} {c}^{j}_{m} {{c^*}^{(N)}}^{j}_{m} },
\end{equation}

\noindent which means that the spectrum of cut  operators
converges towards the spectrum of operators in infinite Hilbert
space. Moreover one can tell how fast is the convergence because
from (49) it is clear that the convergence  $E^{(N)}_m
\xrightarrow[N \longrightarrow \infty]{} E_m$ is governed by the
behavior of the $c^{j}_m$ at large $j$. Note that in (49) $c^{j}_m
$ are the exact components of eigenvectors of $H$. This is exactly
the result we were anticipating because the difference between
localized and non-localized states lies in components $c^{j}_m $.
Therefore one can numerically judge weather the state is bound or
not on grounds of the behavior of the eigenvalues of cut operators
only.

\noindent For the case of a free particle one can obtain
$c^{n}_{E}$ exactly

\begin{equation}
c^{n}_{E}=\langle n \mid  k \rangle = \int_{R} dx \langle n \mid  x \rangle \langle x \mid  k \rangle =
 \int_{R} dx \psi^{H.O.}_n(x) e^{ikx},
\end{equation}

\noindent where H.O. stands for harmonic oscillator

\begin{equation}
\psi^{H.O.}_n(x)=\frac{1}{\sqrt{2^n n! \pi}}H_n(x) e^{-x^2/2}.
\end{equation}

 \noindent Integral (50) is evaluated with the aid of some
analytic properties of Hermite polynomials, what is presented in
Appendix C. The result is \footnote{ Eq. (52) can be obtained
independently in a shorter way. Notice that $c^{n}_{E}$ is a
Fourier transform of $\psi^{H.O.}_n(x)$ which is the solution for
hamiltonian $H=\frac{1}{2}p^{2}+\frac{1}{2}x^{2}$. The Fourier
transformation switches x with p but H is symmetric in those
variables so the Shr\"{o}dinger equation in momentum
representation is the same as in coordinate representation.
Therefore the solution for harmonic oscillator in momentum
representation is of the same form (up to a multiplication factor)
as the solution for harmonic oscillator in coordinate
representation. The connection between those two solutions is
given by Fourier transform hence coefficients $c^{n}_{E}$ are of a form (52) .\\
}

\begin{equation}
c^{n}_{E}=\sqrt{2\pi} i^n \psi^{H.O.}_n(k).
\end{equation}

\noindent Figure 5 is an example of (52) for
$E=\frac{k^2}{2}=1000$.

\begin{figure}[h]
\centering \leavevmode \epsfverbosetrue \epsffile{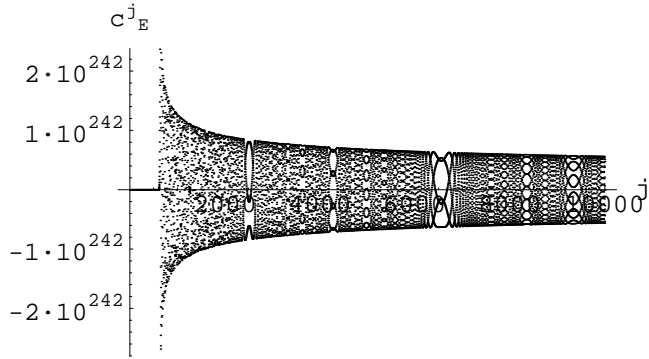}
\caption{Components of the eigenvector (E=1000) for free particle.
}
\end{figure}

\vspace{1cm}

\noindent  Asymptotic behavior of the  envelope is ( see Appendix
C )  $\mid c^{n}_{E} \mid \approx \sqrt[4]{\frac{2}{\pi n}} $
which is indeed power like .

\noindent Similar calculations for discrete  spectrum are not
known, so one is left with numerical data instead. Figure 6
presents components of eigenvector corresponding to the first (the
lowest) eigenvalue of anharmonic oscillator, as well as the
convergence of the first eigenvalue.

\begin{figure}[h]
\centering \leavevmode \epsfverbosetrue \epsffile{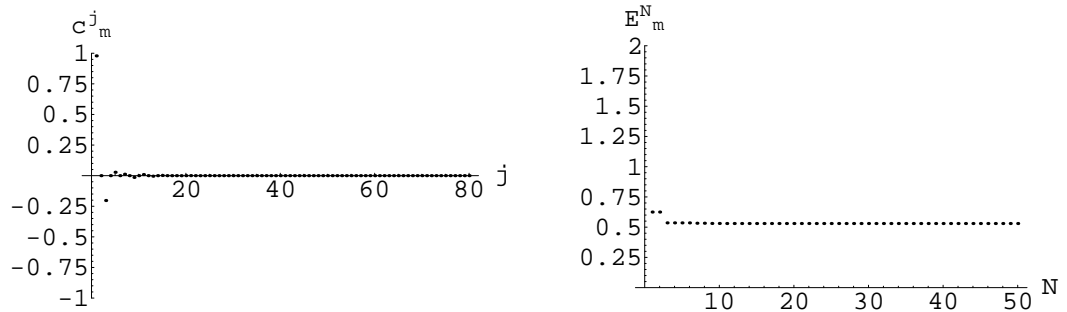}
\caption{Components of the eigenvector  ($c^{j}_{m=1}$) and the
the convergence of eigenvalues ($E^{N}_{m=1}$)  for anharmonic
oscillator. }
\end{figure}

\noindent In this case the  behavior of $c^{j}_{m}$ is completely
different from one shown in Figure 5. One sees that $E^{N}_{m}$
varies in the same ( exponential ) way as $c^{j}_{m}$. In other
words, the behavior of eigenvalues $E^{N}_{m}$ with the cutoff $N$
distinguishes wether the state is bound or not.

\section{Conclusions}

 The main purpose of this paper was to prove that the method
proposed in [7,8]
 enables to distinguish numerically weather the state is localized or
 not. This aim and related problems have already been investigated [9-13].
 This distinction is an important issue while studying supersymmetric models (D=10 SYMQM)
 where bound states exist among dense number of scattering ones [1].
 Therefore one has to reanalyze quantum systems from the very beginning in a new manner.
 Starting from the calculation of spectrum of cut operators
 $Q^{(N)},P^{(N)}$one realizes that eigenvalues of those operators are exactly equal
  to the roots of Hermite polynomials.
 Next, we conclude that in order to recover the continuum limit one has to introduce scaling $m(N)$.
 The validity of the scaling law in the hamiltonian of a free particle was rigourously
 proven in section 5 and numerically tested in section 6. As a result one reproduces the
 dispersion relation from an information about a spectrum of a cut hamiltonian.
 It is expected that the same scaling may be applied for a set of hamiltonians commuting
 with $P$ or under weaker assumptions, namely those for which $P$ can be defined asymptotically.  The scaling in higher dimensions
 is  important because of the occurrence of scattering states (e.g  SYMQM D=2 systems). The formula  (18) is expected to be valid in
 those cases because they are described by quantum mechanics of a free particle in color dimensions.
 In this case the coefficient  in (17) may be different, however (18) is claimed to be applicable
  all the time. In particular  D=2, SU(2) SYMQM [10] is free and it has been found [14] that the system requires (17)
   exactly to recover the continuum limit. Recently a new possibility to speed up the numerical approach in D=4
   has occurred [11]. The naive diagonalization of the Hamiltonian in the whole cut
Hilbert space was abandoned and replaced by the language
   of rotational invariance. The new approach can be extended to higher dimensions as well.
    \\

\section{Acknowledgments}
I am very grateful to my supervisor Prof. Jacek Wosiek for
priceless advices and comments concerning this paper. This work
was supported by the Polish Committee for Scientific Research
under grants no. PB 2P03B 09622 and no. PB 1P03B 02427.

\section{Appendix A}

Here we derive the formula (49). Let us start with eigen equation
$H c_m =E_m c_m$ where $H$ is an operator and $c_m$ its
eigenvector. Writing it in the matrix form

\begin{equation}
 \left[ \begin{array}{ccccccc}
 &             &           &  \multicolumn{1}{|c}{h_{1 \ N+1}}  & \ldots    &             \\
 &   H^{(N)}   &           &  \multicolumn{1}{|c}{\vdots}  & \cdots    &              \\
 &             &           &  \multicolumn{1}{|c}{h_{N \ N+1}}  &  \ldots      &              \\ \cline{1-3}
 h_{N+1 \ 1} & \ldots  & h_{N+1 \ N} &  h_{N+1 \ N+1} & \ldots   &          \\
 \vdots      &  \vdots &  \vdots     &  \vdots        &  \vdots  &
           \end{array}\right]
\left[ \begin{array}{ccccccc}
c^{1}_{m}   \\
\vdots      \\
c^{N}_{m}   \\
c^{N+1}_{m} \\
\vdots      \\
 \end{array}\right] =
E_{m} \left[ \begin{array}{ccccccc}
c^{1}_{m}   \\
\vdots      \\
c^{N}_{m}   \\
c^{N+1}_{m} \\
\vdots      \\
 \end{array}\right],
\end{equation}

\noindent and rewriting  for first N components only one obtains
\begin{equation}
H^{(N)} \left[ \begin{array}{ccccccc}
c^{1}_{m}   \\
\vdots      \\
c^{N}_{m}   \\
\end{array}\right] +
\left[ \begin{array}{ccccccc}
\sum_{i=1}^{\infty} h_{1 \ N+i}c^{N+i}_{m}   \\
\vdots      \\
\sum_{i=1}^{\infty} h_{N \ N+i}c^{N+i}_{m}   \\
 \end{array}\right] =
E_{m} \left[ \begin{array}{ccccccc}
c^{1}_{m}   \\
\vdots      \\
c^{N}_{m}   \\
 \end{array}\right].
\end{equation}

\noindent Now complex conjugate (54) and multiply it by
${c^{(N)}}_n$ from the right side

\[
 \left[ \begin{array}{ccccccc}
{c^{*}}^{1}_{m} & \ldots  &  {c^{*}}^{N}_{m}
\end{array}\right]
H^{(N)}
\left[ \begin{array}{ccccccc}
{c^{(N)}}^{1}_{n}   \\
\vdots      \\
{c^{(N)}}^{N}_{n}   \\
\end{array}\right]
+ \left[ \begin{array}{ccccccc} \sum_{i=1}^{\infty}
{h^*}_{1,N+i}{c^*}^{N+i}_{m} &, \ldots  ,& \sum_{i=1}^{\infty}
{h^*}_{N,N+i}{c^*}^{N+i}_{m}
\end{array}\right]
\left[ \begin{array}{ccccccc}
{c^{(N)}}^{1}_{n}   \\
\vdots      \\
{c^{(N)}}^{N}_{n}   \\
\end{array}\right]
= \]
\begin{equation}
 E_{m} \left[ \begin{array}{ccccccc}
{c^*}^{1}_{m} & \ldots  &   {c^*}^{N}_{m}
 \end{array}\right]
\left[ \begin{array}{ccccccc}
{c^{(N)}}^{1}_{n}   \\
\vdots      \\
{c^{(N)}}^{N}_{n}   \\
\end{array}\right],
\end{equation}

\noindent so that

\begin{equation}
E^{(N)}_{m} \sum_{i=1}^{N} {c^{*}}^{j}_{m} {c^{(N)}}^{j}_{n}  +
\sum_{i=1}^{N} \sum_{j=N+1}^{\infty} {h^*}_{i j} {c^*}^{j}_{n} =
E_{m}\sum_{i=1}^{N} {c^{*}}^{j}_{m} {c^{(N)}}^{j}_{n},
\end{equation}

\noindent or

\begin{equation}
(E_{m}-E^{(N)}_{n}) \sum_{i=1}^{N} {c^{*}}^{j}_{m}
{c^{(N)}}^{j}_{n} = \sum_{i=1}^{N} \sum_{j=N+1}^{\infty} {h^*}_{i
j} {c^*}^{j}_{n},
\end{equation}

\noindent it is non trivial to realize that above equation means
that $\sum_{i=1}^{N} {c^{*}}^{j}_{m} {c^{(N)}}^{j}_{n}
\longrightarrow \delta_{m n }$ thus one can omit the $n$ index and
write

\begin{equation}
 E_m-E^{(N)}_m =
\frac{\sum_{i=1}^{N} \sum_{j=N+1}^{\infty} h_{i j} c^{j}_{m} }{\sum_{i=1}^{N} {c}^{j}_{m} {{c^*}^{(N)}}^{j}_{m} }.
\end{equation}

\noindent Of course this derivation is for the case with discrete
spectrum (discrete index m ) nevertheless for continuous spectrum
the same calculations give

\begin{equation}
(E-E^{(N)}_{n}) \sum_{i=1}^{N} {c^{*}}^{j}_{E} {c^{(N)}}^{j}_{n} =
 \sum_{i=1}^{N} \sum_{j=N+1}^{\infty} {h^*}_{i j} {c^*}^{j}_{E},
\end{equation}

\noindent where
\begin{equation}
H^{(N)} {c^{(N)}}_E =E^{(N)}_E c^{(N)}_E \ \ \ \hbox{and} \ \ \ H
c_E =E c_E  \ \ \  E \in R.
\end{equation}

\noindent This case is discussed in details in sections 6 and 7.

\section{Appendix B}
In this appendix we derive the asymptotic form of the zeros
$q_{m}^{n}$ of the Hermite polynomial $H_{n} (z)$. When $n$ is an
even number they may be obtained using the following relation [14]

\begin{equation}
H_{n} (z) = (-1)^{ \frac{n}{2} } 2^{n} { \frac{1}{2} n }! L_{{ \frac{n}{2} }}^{-\frac{1}{2}} (z^{2}),
\end{equation}

\noindent where $n$ is an even  number, $L_{n}^{\alpha} (z^{2})$
are the generalized Laguerre polynomials with parameter $\alpha$
(in out case $\alpha = -\frac{1}{2}$). Let $  z_{m}^{n} $ ,$
t_{m,\alpha}^{\frac{n}{2}} $ and $   j_{m,\alpha}$ denote the m-th
positive root of   $ H_{n}(z) $, ${L_{\frac{n}{2}}}^{\alpha}(z) $
and $J_{\alpha}(z) $. One has [14]

\begin{equation}  t_{m,\alpha}^{\frac{n}{2}}=
\frac{{j_{m,\alpha}}^{2}}{4k_{\frac{n}{2}}}
\left( 1+\frac{2({\alpha}^{2} - 1)+{j_{m,\alpha}}^{2}}{ 48{k_{\frac{n}{2}}}^{2}} \right)+O(n^{-5})  ,\end{equation}
\noindent where
\begin{equation}  k_{\frac{n}{2},\alpha}=\frac{n}{2}+\frac{\alpha +1}{2}
\ \ \hbox{ and } \ \  (z_{m,\alpha}^{n})^{2} =  t_{m,\alpha}^{\frac{n}{2}} . \end{equation}

\noindent For $\alpha=-\frac{1}{2}$ we obtain $J_{-\frac{1}{2}} (z)=\sqrt{\frac{2}{{\pi}z}}cos(z)$,
therefore $j_{\alpha,m} =\pi(m-\frac{1}{2})$ where $m=1,2\ldots,\ldots, \frac{n}{2}$ and
$k_{\frac{n}{2},\alpha}=\frac{n}{2}+\frac{1}{4} $ so

\[ (z_{m}^{n})^{2}=\frac{{{\pi}^{2}}(m-\frac{1}{2})^{2}}{4(\frac{n}{2}+
\frac{1}{4})} \left( 1+\frac{{{\pi}^{2}}(m-\frac{1}{2})^{2} -\frac{3}{2}}{48(\frac{n}{2}+\frac{1}{4})^{2}} \right)+
O(n^{-5}) =\]

\begin{equation}
= \frac{{{\pi}^{2}}(m-\frac{1}{2})^{2}}{2n+1}
\left( 1+\frac{{{\pi}^{2}}(m-\frac{1}{2})^{2} -\frac{3}{4}}{3(2n+1)^{2}} \right)+O(n^{-5}).
\end{equation}

\noindent Let us define ( $m$ is fixed )

\begin{equation}
f(n):=  \frac{{{\pi}^{2}}(m-\frac{1}{2})^{2}}{2n+1} \left(
1+\frac{{{\pi}^{2}}(m-\frac{1}{2})^{2} -\frac{3}{4}}{3(2n+1)^{2}}
\right) = \frac{a}{n} + \frac{b}{n^2} + \frac{c}{n^3}+ \ldots \ \
,
\end{equation}

\noindent we have

\begin{equation}
z_{m}^{n} = \sqrt{f(n)+O(n^{-5})} = \sqrt{f(n)}
\sqrt{1+\frac{O(n^{-5})}{f(n)}} \approx \sqrt{f(n)}( 1+\frac{1}{2}
O(n^{-4})) = \sqrt{f(n)}(1+ O(n^{-4})),
\end{equation}

\noindent so

\begin{equation}
z_{m}^{n}=   \sqrt{f(n)} + \sqrt{f(n)}  O(n^{-4})=
\sqrt{f(n)}+O(n^{-4.5}),
\end{equation}

\noindent finally

 \begin{equation}
z_{m}^{n}=\frac{\pi(m-\frac{1}{2})}{\sqrt{2n+1}} \sqrt{
1+\frac{{{\pi}^{2}}(m-\frac{1}{2})^{2} -\frac{3}{2}}{3(2n+1)^{2}}
}+O(n^{-4.5}).
\end{equation}

\noindent When $n$ is an odd number there are [10]  analogous relations

\begin{equation}
 H_{n} (z) = (-1)^{ \frac{n-1}{2} } 2^{n} { (\frac{n-1}{2}) }!z L_{{ \frac{n-1}{2} }}^{\frac{1}{2}}
 (z^{2}),
\end{equation}

\noindent and

\begin{equation}
t_{m,\alpha}^{\frac{n-1}{2}}=
\frac{{j_{m,\alpha}}^{2}}{4k_{\frac{n-1}{2}}} \left(
1+\frac{2({\alpha}^{2} - 1)+ {j_{m,\alpha}}^{2}}{
48{k_{\frac{n-1}{2}}}^{2}} \right)+O(n^{-5}), \ \ \end{equation}

\noindent where

\begin{equation}
 k_{\frac{n-1}{2},\alpha}=\frac{n-1}{2}+\frac{\alpha +1}{2}
\ \ \hbox{ and } \ \  (z_{m,\alpha}^{n-1})^{2} =  t_{m,\alpha}^{\frac{n-1}{2}} .
\end{equation}

\noindent  In this  case $ J_{\alpha} (z)=J_{\frac{1}{2}}
(z)=\sqrt{\frac{2}{{\pi}z}}\sin(z) $ so $j_{\alpha,m}=\pi m$ where
$m=1,2\ldots,\ldots, \frac{n-1}{2}$ and
$k_{\frac{n-1}{2},\alpha}=\frac{n}{2}+\frac{1}{4} $. Analogous
calculations give

\begin{equation}
z_{m}^{n}=\frac{\pi m}{\sqrt{2n+1}} \sqrt{ 1+\frac{{{\pi}^{2}}m^{2} -\frac{3}{2}}{3(2n+1)^{2}} }+O(n^{-4.5}).
\end{equation}

\section{Appendix C}

\noindent   Here we evaluate the integral

\begin{equation}
I_n(k)= \int_{R} dx H_n(x) e^{-x^2/2} e^{ikx}.
\end{equation}

\noindent It follows from three properties of Hermite polynomials
[10] that

\begin{equation}
 H_n(x+y)=\frac{1}{2^{n/2}} \sum_{m=0}^{n} \binom{n}{m}
 H_m(\sqrt{2}x)H_{n-m}(\sqrt{2}y),
\end{equation}

\begin{equation}
  H_n(x) = \frac{2^n}{\sqrt{\pi}} \int_{R} dt (x+it)^n e^{-t^2},
\end{equation}

\begin{equation}
  \int_{R} dx H_n(x) H_m(x) e^{-x^2}  = 2^n n! \sqrt{\pi} \delta_{n m}.
\end{equation}

\noindent After substituting (75) to (73) and changing the
variables $x \longrightarrow x+ik$ we get

\begin{equation}
 I_n(k)=\frac{1}{\sqrt{\pi}} 2^{\frac{3n+1}{2}} e^{-k^2/2} i^n \int_{R} dt
e^{-t^2} \int_{R} dx e^{-x^2}
\left(\frac{t+k}{\sqrt{2}}+ix\right)^n.
\end{equation}

\noindent Using (75) once again we obtain

\begin{equation}
 I_n(k)=e^{-k^2/2} 2^{\frac{n+1}{2}} i^n \int_{R} dt e^{-t^2} H_n\left( \frac{t+k}{\sqrt{2}} \right).
\end{equation}

\noindent Finally substituting (74) to (78) and using (76) we get

\begin{equation}
 I_n(k)= e^{-k^2/2} \sqrt{2} i^n \sum_{m=0}^{n} \binom{n}{m} H_{n-m}(k) 2^m m! \sqrt{\pi} \delta_{m 0}=e^{-k^2/2} \sqrt{2\pi} i^n H_n(k),
\end{equation}

\noindent therefore

\begin{equation}
c^{n}_{E}=\langle n \mid  k \rangle =\frac{1}{\sqrt{2^n n! \pi}} I_n(k)=\sqrt{2 \pi} i^n \psi^{H.O.}_n(k).
\end{equation}

 It is straightforward now to estimate components $c^{n}_{E}$.

 \begin{equation}
\mid c^{n}_{E} \mid=\frac{1}{\sqrt{2^n n! \pi}} \mid I_n(k) \mid
\leq \frac{1}{\sqrt{2^n n! \pi}} \mid I_n(0)\mid \leq
\frac{1}{\sqrt{2^n n! \pi}} \sqrt{2 \pi} \mid H_n(0)\mid.
\end{equation}

\noindent Since $H_{2n+1}(0)=0$ hance  $\mid c^{2n+1}_{E} \mid
=0$. On the other hand $H_{2n}(0)={(-1)}^n \frac{(2n)!}{n!}$
therefore $\mid c^{2n}_{E} \mid \leq
\frac{\sqrt{2}\sqrt{(2n)!}}{2^n n!}$ . Finaly according to
Stirling formula one obtains

\begin{equation}
\mid c^{2n}_{E} \mid \lessapprox \sqrt[4]{\frac{2}{\pi n}}.
\end{equation}


\begin{thebibliography}{7}

\bibitem{Banks:phys.rev.}
T. Banks, W. Fischler, S. Shenker and L. Susskind, Phys. Rev.
{\bfseries D55} (1997) 6189; hep-th/9610043. {\itshape }

\bibitem{Witten:nucl.phys.}
E. Witten, Nucl. Phys. {\bfseries B185/188} (1981) 513. {\itshape }


\bibitem{hal:phys.rev.}
M. Claudson and M. B. Halpern, Nucl. Phys. {\bfseries B250} (1985)
689. {\itshape }


\bibitem{hal:phys.rev.}
F. Cooper, A. Khare and  U. Sukhatme  Phys. Rept. {\bfseries 251}
 (1995) 267-385; hep-th/9405029 {\itshape }


\bibitem{hal1:phys.rev.}
S. Samuel, Phys. Lett {\bfseries B411} (1997) 268; hep-th/9705167.
{\itshape }


\bibitem{Coo:susyqm1}
B. de Wit, M. L\"{u}scher and H. Nicolai, Nucl. Phys. {\bfseries
B320} (1989) 135. {\itshape }


\bibitem{wo3:proc.}
J. Wosiek, Supersymmetric Yang-Mills quantum mechanics, in
Proceedings of the NATO Advanced Research Workshop on Confinement,
Topology and Other Non-Perturbative Aspects of QCD, eds. J.
Greensite and S. Olejnik, Kluwer AP, Dordrecht, 2002;
hep-th/0204243. {\itshape }

\bibitem{wo1:nucl.phys}
J. Wosiek, Nucl. Phys. {\bfseries B644} (2002) 85-112;
hep-th/0203116. {\itshape }


\bibitem{trze:acta}
M. Trzetrzelewski and J. Wosiek, Acta Phys. Polon. {\bfseries B35}
(2004) 1615; hep-th/0308007. {\itshape }


\bibitem{wo2:nucl.phys}
M. Campostrini and J. Wosiek, Phys. Lett.  {\bfseries B550} (2002)
121-127; hep-th/0209140. {\itshape }


\bibitem{wo2:nucl.phys}
M. Campostrini and J. Wosiek; hep-th/0407021. {\itshape }

\bibitem{wo1:nucl.phys}
J. Kota\'nski and J. Wosiek, Nucl. Phys. {\bfseries B119} (2003)
932; hep-lat/0208067. {\itshape }

\bibitem{kares:nucl.phys}
V. Kare\v{s} , Nucl. Phys. {\bfseries B689} (2004) 53;
hep-th/0401179. {\itshape }

\bibitem{Bender:Phys. Rev.   }
M. Trzetrzelewski (in preparation) {\itshape }


\bibitem{Abr:Handbook}
M. Abramowitz, I.A.Stegun, {\itshape Handbook of  Mathematical
Functions with Formulas, Graphs, and Mathematical Tables,} Dover
Publications, New York, 1968.



\end{thebibliography}
\end{document}